\documentclass[universe,communication,accept,moreauthors,pdftex]{Definitions/mdpi}
\usepackage{amsmath}
\usepackage[utf8]{inputenc}
\usepackage{latexsym}
\usepackage{amsfonts}
\usepackage{graphicx,color}
\usepackage{mathrsfs}
\usepackage{epstopdf}
\usepackage{subfigure}

\usepackage{booktabs}
\usepackage{multirow}
\usepackage{soul}
\usepackage{microtype}
\graphicspath{{Definitions/}}

\firstpage{1}
\pubvolume{5}
\issuenum{1}
\articlenumber{5}
\pubyear{2020}
\copyrightyear{2020}
\history{Received: date; Accepted: date; Published: date}

\Title{The Reconstruction of Non-Minimal Derivative Coupling Inflationary Potentials}

\Author{Qin Fei $^{1}$, Zhu Yi $^{2,}$*\orcidA{} and Yingjie Yang $^{3}$}

\address{%
$^{1}$ \quad School of Mathematics and Physics, Hubei Polytechnic University, Huangshi 435003, China; feiqin@hbpu.edu.cn\\ 
$^{2}$ \quad Department of Astronomy, Beijing Normal University, Beijing 100087, China\\
$^{3}$ \quad School of Physics, Huazhong University of Science and Technology, Wuhan 430074, China; yyj@hust.edu.cn} 
\corres{Correspondence: yz@bnu.edu.cn}

\abstract{We derive the reconstruction formulae for the inflation model with the non-minimal derivative coupling term. If reconstructing the potential from the tensor-to-scalar ratio $r$, we~could obtain the potential without using the high friction limit. As an example, we~reconstruct the potential from the parameterization $r=8\alpha/(N+\beta)^{\gamma}$, which is a general form of the $\alpha$-attractor. The~reconstructed potential has the same asymptotic behavior as the T- and E-model if we choose $\gamma=2$ and $\alpha\ll1$. We~also discuss the constraints from the reheating phase by assuming
the parameter $w_{re}$ of state equation during reheating is a constant. The~scale of big-bang nucleosynthesis could put an upper  limit on $n_s$ if $w_{re}=2/3$ and a low limit on $n_s$ if $w_{re}=1/6$.}
\begin{document}

\section{Introduction}
In the standard big-bang cosmology, inflation has successfully solved various problems,
such as the flatness, horizon and monopole problems. Besides, its quantum
fluctuation can produce the seed of the formation of large-scale structure~\cite{Starobinsky:1980te, Guth:1980zm, Linde:1983gd, Albrecht:1982wi}.
A scalar field with a flat potential is usually chosen to investigate inflation. The~most economical and fundamental candidate for the inflaton is therefore the Standard Model Higgs boson. However, the Higgs boson is disfavored by the observational data~\cite{Linde:1983gd,Akrami:2018odb} when minimally coupled to gravity due to its large tensor-to-scalar ratio.
If the kinetic term of the scalar field is non-minimal coupled to Einstein tensor, the tensor-to-scalar ratio $r$ could be reduced to being consistent with the observational data, and~the effective Higgs self-coupling $\lambda$ could be the order of $1$~\cite{Germani:2010gm,Germani:2014hqa}.
This inflation model with non-minimal derivative coupling belongs to the subclass of the Horndeski theory~\cite{Horndeski:1974wa}, which is a general scalar--tensor theory, with field equations that are at most of the second-order derivatives of both the metric $g_{\mu\nu}$ and scalar field $\phi$ in four dimensions~\cite{Sushkov:2009hk}. Therefore, the non-minimal derivative coupling inflation model could save the Higgs model without introducing a new degree of freedom. For more about the non-minimal derivative coupling inflation model, refer to~\cite{Yang:2015pga,Yang:2015zgh,Huang:2015yva,Gong:2017kim,Fu:2019ttf,Oikonomou:2020sij,Odintsov:2020sqy,Gialamas:2020vto}.

The most important observables of inflation are the spectral tilt $n_s$ and the tensor-to-scalar ratio $r$. To be compared with the observational data easily, they are usually expressed by the $e$-folding number $N$ before the end of inflation at the horizon exit of the pivotal scale. Among them, one of the predictions that is greatly favored by the observational data may be the $\alpha$-attractors, $n_s=1-2/N$ and $r=12\alpha/N^2$.  Numerous inflation models make the $\alpha$-attractors prediction, for example  the Starobinsky model~\cite{Starobinsky:1980te}, the Higgs inflation with a non-minimal coupling $\xi\phi^2 R$ in the strong coupling limit $\xi\gg1$~\cite{Kaiser:1994vs,Bezrukov:2007ep}, the pole inflation with the kinetic term being $(\partial\phi)^2/(1-\phi^2/6\alpha)^2$~\cite{Kallosh:2013yoa} and the T/E model~\cite{Kallosh:2013hoa,Kallosh:2013maa}. It is therefore worth studying  whether  there are still other models that can make the prediction of $\alpha$-attractors. In this paper, we~consider the non-minimal derivative coupling inflation models to investigate this $\alpha$-attractors issue by reconstructing the potential.
Starting from the observational data and parameterizing the observable with $N$, using the relationships between the observable and the potential, we~can then reconstruct the potential~\cite{Huang:2007qz, Gobbetti:2015cya}. By this reconstruction, the model parameters can be constrained easily and the reconstructed potential would   always be  consistent with the observational data\cite{Mukhanov:2013tua,Roest:2013fha,Garcia-Bellido:2014gna,
Garcia-Bellido:2014wfa,Garcia-Bellido:2014eva,Creminelli:2014nqa,
Boubekeur:2014xva,
Barranco:2014ira,Galante:2014ifa,Gobbetti:2015cya,Chiba:2015zpa,
Cicciarella:2016dnv,Lin:2015fqa,Nojiri:2010wj,
Odintsov:2016vzz,Yi:2016jqr,Odintsov:2017qpp,Nojiri:2017qvx,
Choudhury:2017cos,Gao:2017uja,Jinno:2017jxc,Gao:2017owg,Fei:2017fub,Koh:2016abf}.

After the inflation, it is followed by the reheating phase, which may give additional constraints on the inflation phase~\cite{Dai:2014jja,Fei:2017fub}.
Assuming that the effective parameter $w_{re}$ of state equation during reheating is a constant and the entropy is a conserved quantity, we~can
relate the $e$-folding number and the energy scale during reheating to those during inflation~\cite{Dai:2014jja,Cook:2015vqa,Ueno:2016dim,Kabir:2016kdh,
DiMarco:2017sqo,Dimopoulos:2017zvq,Gong:2015qha}. From these relations, the constraints on the energy scale during reheating would transfer to the constraints on the inflation~model.

In this paper, we~reconstruct the inflationary potentials of the non-minimal coupling inflation models and research the additional constraints from the reheating phase. The~paper is organized as follows. In Section~\ref{sec-2}, we~give a brief review about the inflation model with the non-minimal derivative coupling term and the reconstruction method. In Section~\ref{sec-3}, we~reconstruct the potential from the parameterization of tensor-to-scalar ratio $r$. We discuss the constraints from the reheating in Section~\ref{sec-4}, and~give the conclusion in Section~\ref{sec-5}.

\section{The Relations}
\label{sec-2}
In this section, we~develop the formulae for the reconstruction of the inflationary potential with the kinetic term non-minimal coupled to Einstein tensor. We start from the action
\begin{equation}\label{action}
\begin{aligned}
 S=\frac{1}{2}\int d^4 x\sqrt{-g}\bigg[R-g^{\mu\nu}\partial_\mu\phi\partial_\nu\phi
 +\frac{1}{M^2}G^{\mu\nu} \partial_\mu\phi\partial_\nu\phi-2V(\phi)\bigg],
\end{aligned}
\end{equation}
where we choose the unit $c=M_{pl}^2=1/(8\pi G)=1$ and the coupling parameter $M$ is a constant with the dimension of mass.
For the homogeneous and isotropic Universe with the Robertson--Walker~metric
\begin{equation}
ds^2=-dt^2+a(t)^2\left[\frac{dr^2}{1-Kr^2}+r^2\left(d\theta^2+\sin^2\theta d\phi^2\right)\right],
\end{equation}
where $K=0$ in the inflation epoch, the action \eqref{action} becomes
\begin{equation}
S=\frac{1}{2}\int d^4 x\sqrt{-g}\bigg[R+\left(1+ \frac{3 H^2}{M^2}\right)\dot{\phi}^2-2V(\phi)\bigg].
\end{equation}
The kinetic term of this model is 
\begin{equation}
 (1+ \frac{3 H^2}{M^2})\frac{\dot{\phi}^2}{2}>0,
\end{equation}
so there are  no  ghosts in this model. The~scale range of the parameter $M$ is very broad. If $M$ is extremely larger than the Hubble parameter, $M^2\gg H^2$, the non-minimal derivative coupling term can be neglected and the model reduces to the canonical case. If $M$ is extremely smaller than the Hubble parameter, $M^2\ll H^2$, the non-minimal derivative coupling term   dominates  the inflation, and~may make some new predictions different from the canonical case.

The Friedmann equation is
\begin{equation}\label{eq1}
 H^2=\left(\frac{\dot{a}}{a}\right)^2=\frac{1}{3}\left[\frac{\dot{\phi}^2}{2}\left(1+9 F\right)+V(\phi)\right],
\end{equation}
where $F=H^2/M^2$ is the friction parameter. The~equation of motion for the scalar field $\phi$ is
\begin{equation}\label{eq2}
 \frac{d}{dt}\left[a^3\dot{\phi}(1+3F)\right]=-a^3\frac{dV}{d\phi}.
\end{equation}
For the slow-roll inflation, the slow-roll conditions are 
\begin{align}
\label{slrcond1}
\begin{split}
\frac{1}{2}(1+9F)\dot\phi^2 &\ll V(\phi),\\
\big|\ddot \phi\big| &\ll \big|3H\dot\phi\big|,\\
\bigg|\frac{2\dot H}{M^2+3H^2}\bigg| &\ll 1.
\end{split}
\end{align}
Under these slow-roll conditions, the background Equations~\eqref{eq1}~and~\eqref{eq2} become 
\begin{gather}
\label{eq4}
H^2 \approx \frac{V(\phi)}{3},\\
\label{eq5}
3 H\dot\phi(1+3F) \approx - V_\phi,
\end{gather}
where $V_\phi=dV/d\phi$. With Equation~\eqref{eq4}, the friction parameter becomes
\begin{equation}\label{sf}
 F\approx \frac{V(\phi)}{3M^2}.
\end{equation}
 The corresponding slow-roll parameters are
\begin{gather}
 \label{sl1}
 \epsilon_V=\frac{1}{2}\left(\frac{V_\phi}{V}\right)^2\frac{1+9F}{(1+3F)^2},\\
 \label{sl2}
 \eta_V=\frac{1}{1+3F}\frac{V_{\phi\phi}}{V}.
\end{gather}
Using Equations~\eqref{eq4}, \eqref{eq5}~and~\eqref{sl1}, we~obtain
\begin{equation}\label{phiV}
 \frac{3\dot{\phi}^2(1+9F)}{2V(\phi)}\approx \epsilon_V.
\end{equation}
The derivative of $\epsilon_V$ with respect to $t$ is~\cite{Yang:2015pga}
\begin{align} \label{depsilon}
 \dot{\epsilon_V}&=2H\epsilon_V\bigg[\frac{2+21F+81F^2}{(1+9F)^2}
 \epsilon_V-\eta_V-\frac13\eta_V^2\notag\\
 &-\frac{4+72F+603F^2+2538F^3+5103F^4}{3(1+3F)(1+9F)^3}\epsilon_V^2\notag\\
&+\frac{2(2+48F+441F^2+1944F^3+3645F^4)}{3(1+3F)(1+9F)^3}\epsilon_V\eta_V\bigg].
\end{align}
By using the relation $dN=-Hdt$, to the first order of slow-roll parameters, Equation~\eqref{depsilon} becomes
\begin{equation}\label{eps-eta}
 \frac{d\ln\epsilon_V}{dN}=2\left[\eta_V-\frac{2+21F+81F^2}{(1+9F)^2}
 \epsilon_V\right],
\end{equation}
where $N$ is the e-folding number before the end of inflation at the horizon exit. The~power spectrum for the scalar perturbation is~\cite{Yang:2015pga}
\begin{equation}\label{sca:per}
 P_\zeta\approx \frac{1+9F}{1+3F} \times\frac{H^2}{8\pi^2\epsilon_V}.
\end{equation}
The power spectrum for the tensor perturbation is~\cite{Yang:2015pga}
\begin{equation}\label{ten:per}
 P_T\approx \frac{2H^2}{\pi^2}.
\end{equation}
The scalar tilt $n_s$ and the tensor-to-scalar ratio $r$ are~\cite{Tsujikawa:2012mk,Yang:2015pga}
\begin{gather}
 n_s-1 = 2\eta_V-\frac{6(1+4F)}{1+9F}\epsilon_V, \label{ns}\\
 r =\frac{16(1+3F)}{1+9F}\epsilon_V \label{r}.
\end{gather}
From Equations~\eqref{eps-eta}~and~\eqref{ns}, we~obtain the relation between $n_s$ and $\epsilon_V$,
\begin{equation}\label{epsilonVns}
 n_s-1=\frac{d\ln\epsilon_V}{d N}-\frac{2+36F+54F^2}{(1+9F)^2}\epsilon_V.
\end{equation}
From Equations~\eqref{eq1}~and~\eqref{phiV}, we~obtain the relation between $\phi$ and $N$,
\begin{equation}\label{dNphi}
 d\phi=\pm\sqrt{\frac{2\epsilon_V}{1+9F}}dN,
\end{equation}
where the sign $\pm$ depends on the sign of $dV/d\phi$. Without loss of generality, in this paper, we~only research the `$+$' case.
Combining Equations~\eqref{sl1}~and~\eqref{dNphi}, we~get the relation between the potential and the slow-roll parameter,
\begin{equation}\label{epsilonV}
 \epsilon_V=\frac{1+9F}{2+6F}(\ln V)_{,N}.
\end{equation}
By using Equations~\eqref{sf}~and~\eqref{r},   Equations~\eqref{sca:per}, \eqref{epsilonVns}~and~\eqref{epsilonV} become
\begin{align}
\label{sca:per:r}
 P_\zeta&= \frac{2H^2}{\pi^2r},\\
\label{nsr}
n_s-1 & =\frac{d\ln r}{dN}-\frac{r}{8}, \\
\label{rV}
r&=8(\ln V)_{,N}.
\end{align}
These relations \eqref{sca:per:r}, \eqref{nsr}~and~\eqref{rV} do not contain the friction parameter $F$,  thus  it is possible to reconstruct the potential from the tensor-to-scalar ratio without using the high friction limit. In the following sections, we~  discuss this issue.

\section{The Reconstruction}
\label{sec-3}
In this section, we~  reconstruct the potential from the tensor-to-scalar ratio $r$. The~observational data favor small $r$, and
the $\alpha$-attractor gives $r=12\alpha/N^2$, which is small enough to be consistent with the observational data when $\alpha\ll1$. In this section, we~discuss a general parameterization of the $\alpha$-attractor
\begin{equation}\label{parar}
 r=\frac{8 \alpha}{(N+\beta)^\gamma},
\end{equation}
where $\gamma>1$, and~$\beta$ accounts for the contribution from the scalar field $\phi_e$ at the end of the inflation. From the relation \eqref{nsr}, we~obtain the spectral tilt
\begin{equation}\label{parar4}
 n_s-1=-\frac{\gamma}{N+\beta}-\frac{\alpha}{(N+\beta)^\gamma}.
\end{equation}
With the help of relation \eqref{rV}, we~obtain the potential
\begin{equation}\label{parar1}
 V=V_0\exp\left[-\frac{\alpha}{(\gamma-1)(N+\beta)^{\gamma-1}}\right].
\end{equation}
Combining the slow-roll Friedmann Equation~\eqref{eq4} and the power spectrum in Equation~\eqref{sca:per:r}, we~relate the amplitude of the power spectrum $A_s$ to the potential,
\begin{equation}\label{rel:amp:p}
A_s=\frac{2V}{3\pi^2 r}.
\end{equation}
Substituting the reconstructed potential \eqref{parar1} into relation \eqref{rel:amp:p} and using the parameterization \eqref{parar}, we~obtain
\begin{equation}\label{parar2}
 V_0=\frac32\pi^2 A_s r\exp\left[\frac{\alpha}{\gamma-1}\left(\frac{r}{8\alpha }\right)^\frac{\gamma-1}{\gamma}\right].
\end{equation}
Combining Equation~\eqref{parar1} and   \eqref{epsilonV}, we~get the slow-roll parameter
\begin{equation}\label{parar3}
 \epsilon_V=\frac{1+3F_0\exp\left[\alpha(1-\gamma)^{-1}(N+\beta)^{1-\gamma}\right]}
 {2+2F_0\exp\left[\alpha(1-\gamma)^{-1}(N+\beta)^{1-\gamma}\right]} \frac{\alpha}{(N+\beta)^\gamma},
\end{equation}
where the amplitude of the friction parameter $F_0=V_0/M^2$. From the condition of the end of inflation, $\epsilon_V(0)=1$, we~obtain the relation among $\alpha$, $\beta$ and $\gamma$
\begin{equation}\label{pararp}
 \frac{1+3F_0\exp\left[\alpha(1-\gamma)^{-1}\beta^{1-\gamma}\right]}{2+2F_0
 \exp\left[\alpha(1-\gamma)^{-1}\beta^{1-\gamma}\right]}\times\frac{\alpha}{\beta^\gamma}=1.
\end{equation}
Under the GR limit $F_0\ll1$, relation \eqref{pararp} reduces to $\alpha=2\beta^\gamma$; under the high friction limit $F_0\gg1$, relation \eqref{pararp} reduces to $\alpha=2\beta^\gamma/3$. From Equation~\eqref{parar}, the tensor-to-scalar ratio $r$ under the high friction limit is therefore smaller than that under the GR limit when $\beta$ and $\gamma$ is unchanged. Substituting Equation~\eqref{parar3} into Equation~\eqref{dNphi}, we~get the relation between $\phi$ and $N$,
\begin{align}\label{parar5}
 d\phi=\sqrt{r\Bigg(8+8F_0\exp\Big[ \frac{\alpha(N+\beta)^{1-\gamma}}{(1-\gamma)}\Big]\Bigg)^{-1}}d N.
\end{align}
Combining it with Equation~\eqref{parar}, the relation becomes
\begin{equation}\label{parar51}
d\phi=\sqrt{r\Bigg(8+8F_0\exp\Big[ \frac{\alpha (8\alpha/r)^{(1-\gamma)/\gamma} }{(1-\gamma)}\Big]\Bigg)^{-1}}d N.
\end{equation}
To the first order of tensor-to-scalar ratio $r$, it becomes
\begin{equation}\label{parar6}
 d\phi=\sqrt{\frac{r}{8+8F_0}}d N,
\end{equation}
and the solution is
\begin{equation}\label{parar7}
\phi-\phi_0=\left\{
\begin{aligned}
& \frac{2}{2-\gamma}\sqrt{\frac{\alpha }{1+F_0}}(N+\beta)^{\frac{2-\gamma}{2}},\quad \gamma\neq2,\\
&\sqrt{\frac{\alpha }{1+F_0}}\ln(N+\beta), \quad \gamma=2,
\end{aligned}
\right.
\end{equation}
where $\phi_0$ is the integration constant.
Substituting Equation~\eqref{parar7} into Equation~\eqref{parar1}, we~get the reconstructed potential
\begin{equation}\label{pararv}
V(\phi)=\left\{
\begin{aligned}
&V_0\exp\left[-\lambda\left(\sqrt{1+F_0}\phi_0-\sqrt{1+F_0}\phi\right)^{\frac{2\gamma-2}{\gamma- 2}}\right],\quad \gamma\neq2,\\
&V_0\exp\left[-\alpha e^{-\sqrt{1+F_0}(\phi-\phi_0)/\sqrt{\alpha}}\right], \quad \gamma=2,
\end{aligned}
\right.
\end{equation}
where
\begin{align}\label{parar8}
 \lambda=\frac{\alpha}{\gamma-1}\left(\frac{\gamma-2}{2\sqrt{\alpha}}\right)^{\frac{2\gamma-2}{\gamma-2}}.
\end{align}
Therefore, we~reconstruct the potential from the parameterization \eqref{parar} without using the high friction limit. Furthermore, the potential \eqref{pararv} and parameter \eqref{parar8} show that the effect of the no-minimally derivative coupling term is the rescaling of the inflaton field by a factor $\sqrt{1+F_0}$.
For the $\alpha$-attractors parameterization $\gamma=2$, under the GR limit $F_0\ll1$, the potential reduces to\cite{Yi:2016jqr}
\begin{equation}\label{parar2v}
 V(\phi)=V_0\exp\left[-\alpha e^{-(\phi-\phi_0)/\sqrt{\alpha}}\right].
\end{equation}
If $\alpha\ll1$, this potential reduces to
\begin{equation}
 V(\phi)=V_0 \left[1-\alpha e^{-(\phi-\phi_0)/\sqrt{\alpha}}\right],
\end{equation}
which is asymptotic behavior of the T-model and E-model.

\begin{figure}[H]
 \centering
 \includegraphics[width=0.45\textwidth]{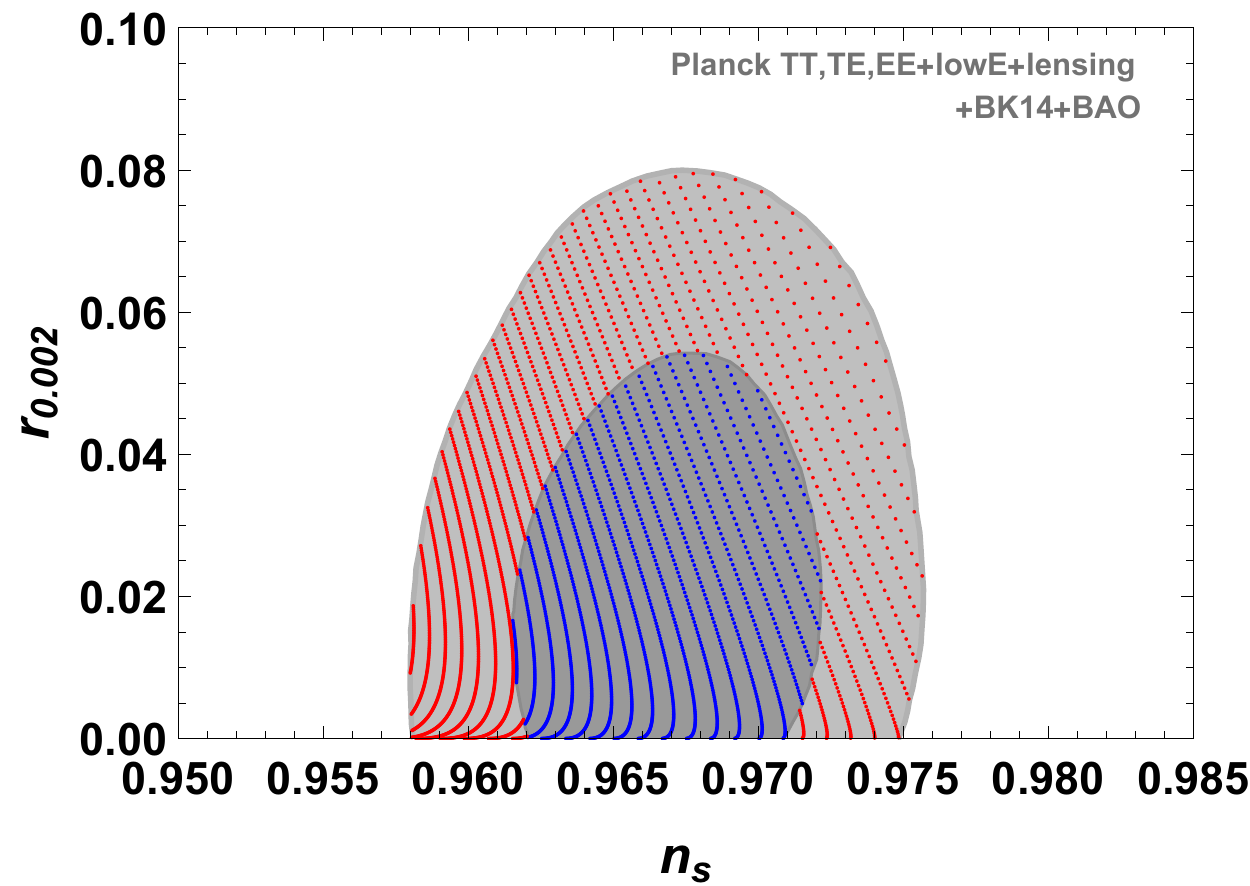}
 \includegraphics[width=0.432\textwidth]{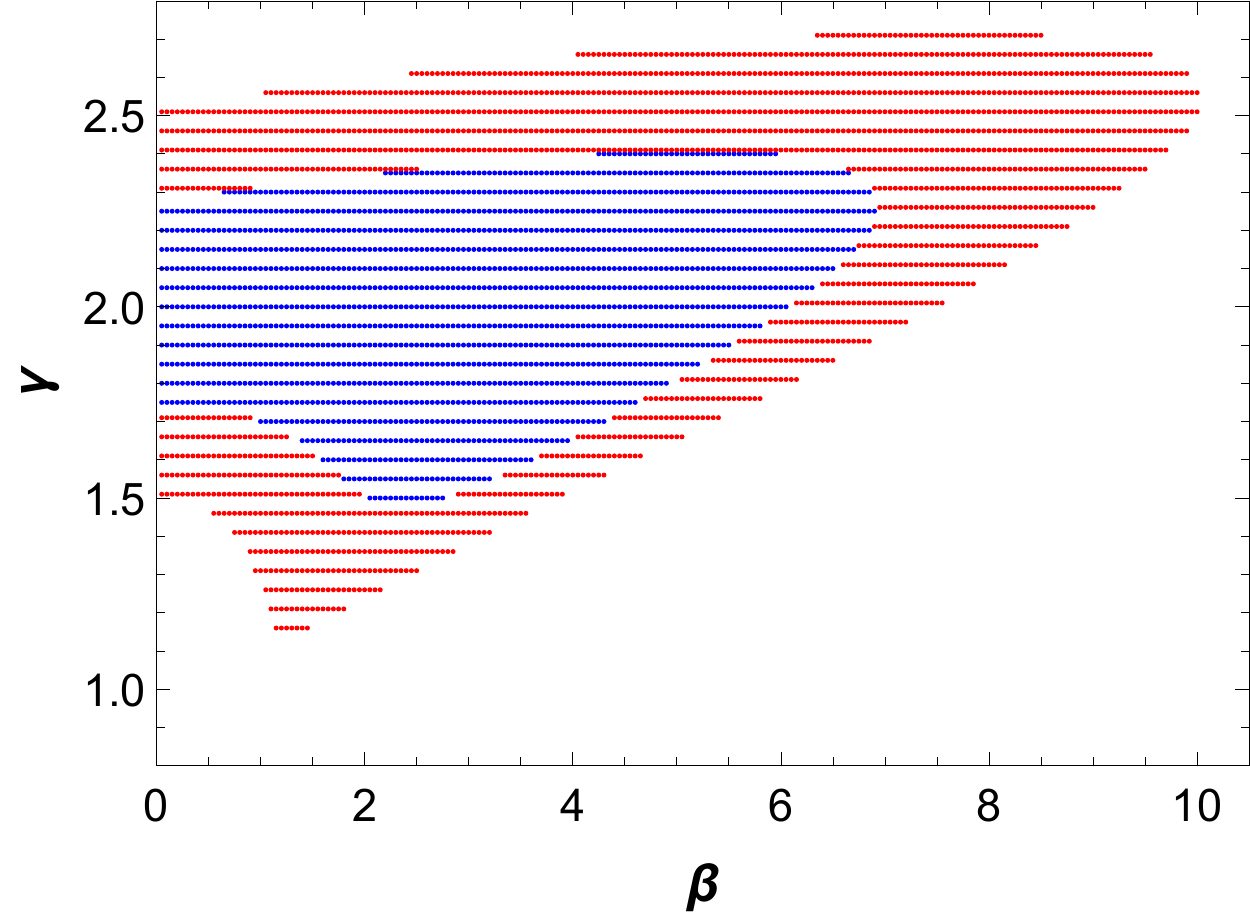}
 \caption{The constraints on $n_s$ and $r_{0.002}$ from Planck data~\cite{Akrami:2018odb} and the theoretical predictions for the parameterization \eqref{parar} in the high friction limit. The~Planck constraints on $n_s$ and $r$ are displayed in the left panel and the constraints on $\beta$ and $\gamma$ for $N=60$ are displayed in the right panel. The~red and blue regions denote the $68\%$ and $95\%$ confidence level, respectively.}\label{pparar1}
\end{figure}

\begin{figure}[H]
 \centering
 \includegraphics[width=0.70\textwidth]{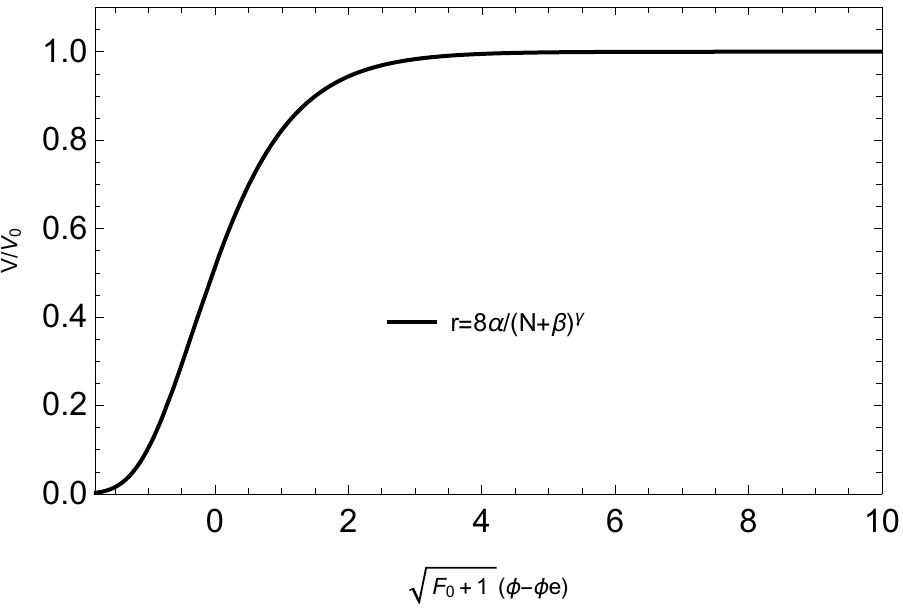}
 \caption{The reconstructed potentials are normalized with $V_0$ from Equation~\eqref{parar2}, and~the inflaton field is normalized with $1/\sqrt{F_0+1}$. We choose the value of $\phi_0$ that could make $\phi_e=0$.}\label{potential}
\end{figure}

Taking $N=60$ and $F_0\gg1$, and comparing the theoretical predictions \eqref{parar}~and~\eqref{parar4} with the Planck 2018 data\cite{Akrami:2018odb}, we~obtain the constraints on the parameters $\beta$ and $\gamma$ shown in Figure~\ref{pparar1}. Taking~$\gamma=2$, $\beta=1$ and $N=60$, the theoretical predictions are $n_s=0.967$ and $r=0.0014$. With these parameters, the plot of the potential is shown in Figure~\ref{potential}. 
\unskip

\section{Reheating}\label{sec-4}
The inflation ends when the inflaton rolls down to the minimum of the potential; around the minimum,
the inflaton field will oscillate to reheat the cold universe. Because the inflation phase is followed by the reheating phase, these two phases may constrain each other, so the reheating phase may give other constraints on the inflation phase. In this section, we~  research the constraint from the reheating phase on the reconstructed model under the high friction limit $F\gg1$ and the GR limit $F\ll1$.

The relation between the pivotal scale $k_\ast=0.002~\text{Mpc}^{-1}$ and the present Hubble parameter is
\begin{equation}
\begin{aligned}
\label{kstar}
\frac{k_\ast}{a_0 H_0}=\frac{a_\ast H_\ast}{a_0 H_0} =\frac{a_\ast}{a_{e}}\frac{a_{e}}{a_{re}}\frac{a_{re}}{a_0}\frac{H_\ast}{H_0} =e^{-N -N_{re}}\frac{a_{re}}{a_0}\frac{H_\ast}{H_0},
\end{aligned}
\end{equation}
where $N_{re}$ is the $e$-folding number during reheating, $a_{re}$ is the scale factor at the end of reheating, and~we assume the radiation domination phase follows the reheating phase immediately and the reheating phase follows inflation phase immediately. Because the physics of the reheating is still unknown, for simplicity, we~assume a constant parameter $w_{re}$ of state equation during reheating, and~we get
\begin{equation}
\label{reheq2}
N_{re}=\frac{1}{3(1+w_{re})}\ln\frac{\rho_{e}}{\rho_{re}},
\end{equation}
where the relation between $\rho_{re}$ and the temperature $T_{re}$ is
\begin{equation}
\label{reheq01}
\rho_{re}=\frac{\pi^2}{30}g_{re}T^4_{re},
\end{equation}
with $g_{re}$ denoting the effective number of relativistic species at reheating phase. By using the condition of the entropy conservation, we~get the relation between temperature $T_{re}$ and the present cosmic microwave background temperature $T_0=2.725K$, 
\begin{equation}
\label{reheq3}
a_{re}^3 g_{s,re}T_{re}^3=a_0^3\left(2 T_0^3+6\times\frac78 T_{\nu0}^3\right),
\end{equation}
where $g_{s,re}$ denotes the effective number of relativistic species for entropy, and~$T_{\nu0}=(4/11)^{1/3}T_0$ is the present neutrino temperature.
By using the above relations, we~obtain~\cite{Dai:2014jja,Cook:2015vqa}
\begin{gather}
\label{Nre}
N_{re}=\frac{4}{1-3w_{re}}\bigg[-N -\ln\frac{\rho_{e}^{1/4}}{H_\ast}+\frac{1}{3}\ln\frac{43}{11g_{s,re}}
+\frac14\ln\frac{\pi^2 g_{re}}{30}-\ln\frac{k_{\ast}}{a_0T_0}\bigg],\\
\label{Tre}
T_{re}=\exp\left[-\frac{3N_{re}(1+w_{re})}{4}\right]\left[\frac{30\rho_{e}}{\pi^2 g_{re}}\right]^{1/4}.
\end{gather}
The relations \eqref{Nre}~and~\eqref{Tre} show that $N_{re}$ and $T_{re}$ depend on $g_{re}$ and $g_{s,re}$ logarithmically,  thus  we choose $g_{re}=g_{s,re}=106.75$. At the end of inflation, we~have $\epsilon_V\approx 1$; from Equation~\eqref{phiV}, we~obtain the relation $\dot{\phi}^2=2V_e/(27F)$, so we have $\rho_e=4V_e/3$. By using the observational value of the amplitude of the power spectrum~\cite{Akrami:2018odb}, from Equation~\eqref{sca:per}, we~have
\begin{equation}\label{re:as}
 A_s=3H_*^2/(8\pi^2\epsilon_{V*})=2.2\times 10^{-9},
\end{equation}
and Equations~\eqref{Nre}~and~\eqref{Tre} become
\begin{gather}
\label{Nre:final}
N_{re}=\frac{4}{1-3w_{re}}\left(56.46-N-\frac{\ln V_{e}}{4}+\frac{\ln\epsilon_{V*}}{2}\right),\\
\label{reheq5}
T_{re}=\exp\left[-\frac{3N_{re}(1+w_{re})}{4}\right]\left[\frac{4V_{e}}{10.675 \pi^2}\right]^{1/4}.
\end{gather}
By using Equations~\eqref{parar1}~and~\eqref{parar3}, under the high friction limit $F\gg1$, we~obtain the constraint from the reheating process on the model parameters,
\begin{equation}
\label{Nre:parar2}
\begin{aligned}
N_{re}=\frac{4}{1-3w_{re}}\bigg[60.45+ \frac{\alpha}{4(\gamma-1)\beta^{\gamma -1}}+\frac14\ln\alpha
 -N-\frac{\gamma}{4}\ln(N+\beta)-\frac{\alpha}{4(\gamma-1)(N+\beta)^{\gamma-1}}\bigg],\\
\end{aligned}
\end{equation}
\begin{equation}
\label{Tre:parar2}
\begin{aligned}
T_{re}=\,\,0.01\,\frac{\alpha^{1/4}}{(N+\beta)^{\gamma/4}}\exp\bigg[ -\frac{\alpha}{4(\gamma-1)\beta^{\gamma-1}}
 +\frac{\alpha}{4(\gamma-1)(N+\beta)^{\gamma-1}}-\frac{3N_{re}(1+w_{re})}{4} \bigg],
\end{aligned}
\end{equation}
where $\alpha=2\beta^\gamma/3$. Under the GR limit $F\ll1$, the relations are
\begin{equation}
\label{Nre:parar:GR2}
\begin{aligned}
N_{re}=\frac{4}{1-3w_{re}}\bigg[59.90+ \frac{\alpha}{4(\gamma-1)\beta^{\gamma -1}}+\frac14\ln\alpha
 -N-\frac{\gamma}{4}\ln(N+\beta)-\frac{\alpha}{4(\gamma-1)(N+\beta)^{\gamma-1}}\bigg],\\
\end{aligned}
\end{equation}
\begin{equation}
\label{Tre:parar:GR2}
\begin{aligned}
T_{re}=\,\,0.01\,\frac{\alpha^{1/4}}{(N+\beta)^{\gamma/4}}\exp\bigg[ -\frac{\alpha}{4(\gamma-1)\beta^{\gamma-1}}
 +\frac{\alpha}{4(\gamma-1)(N+\beta)^{\gamma-1}}-\frac{3N_{re}(1+w_{re})}{4} \bigg],
\end{aligned}
\end{equation}
where $\alpha=2\beta^\gamma$. These two situations make almost the same constraint except the $0.5$ $ e$-folding difference in $N_{re}$ and the different relations of $\alpha$. Therefore, the friction parameter $F$ has little influence on the reheating phase, and~we just consider the high friction limit situation in the following.

\begin{figure}[H]%
\centering
$\begin{array}{cc}
\includegraphics[width=0.45\textwidth]{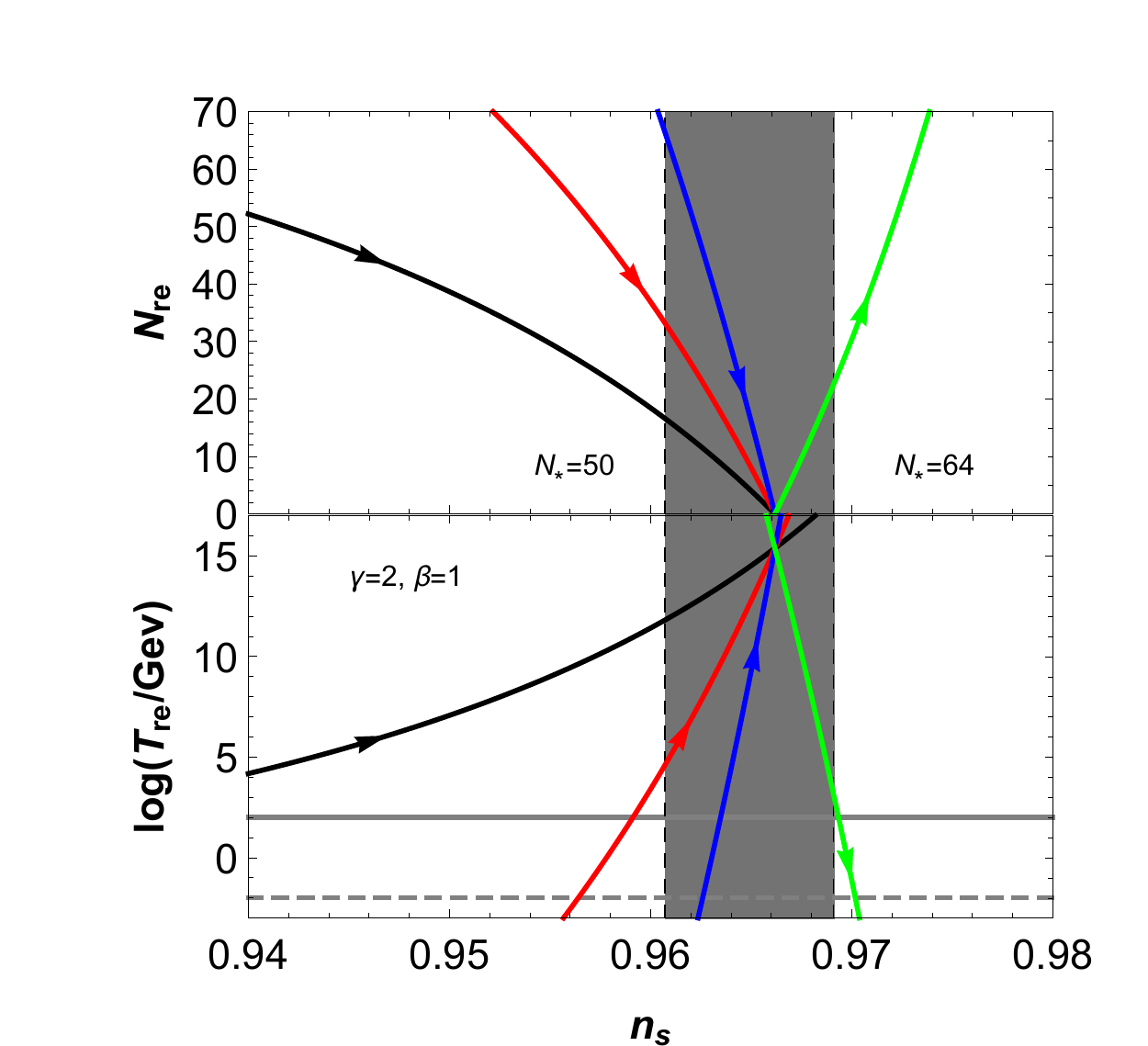}&
\includegraphics[width=0.45\textwidth]{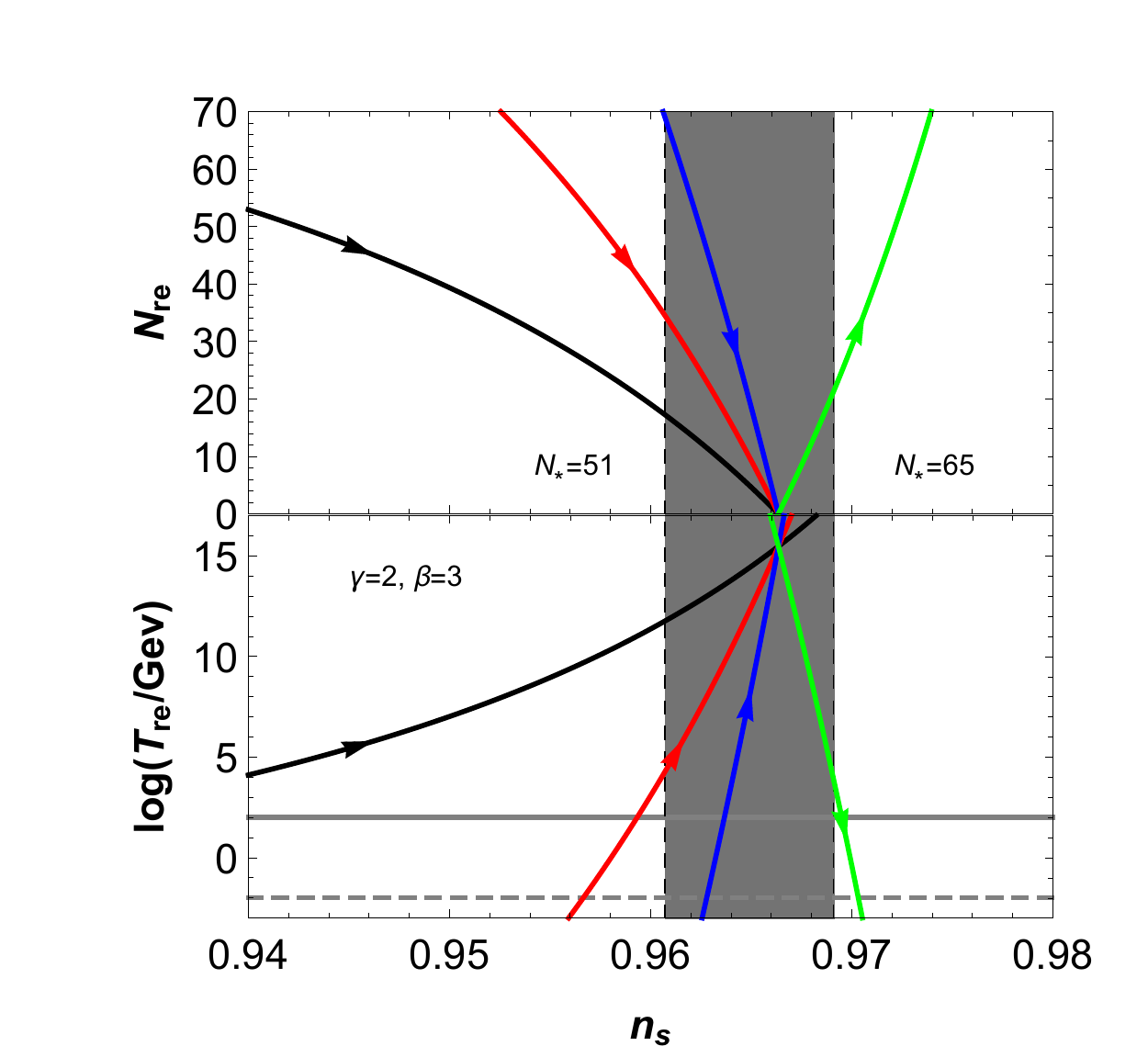}\\
\includegraphics[width=0.45\textwidth]{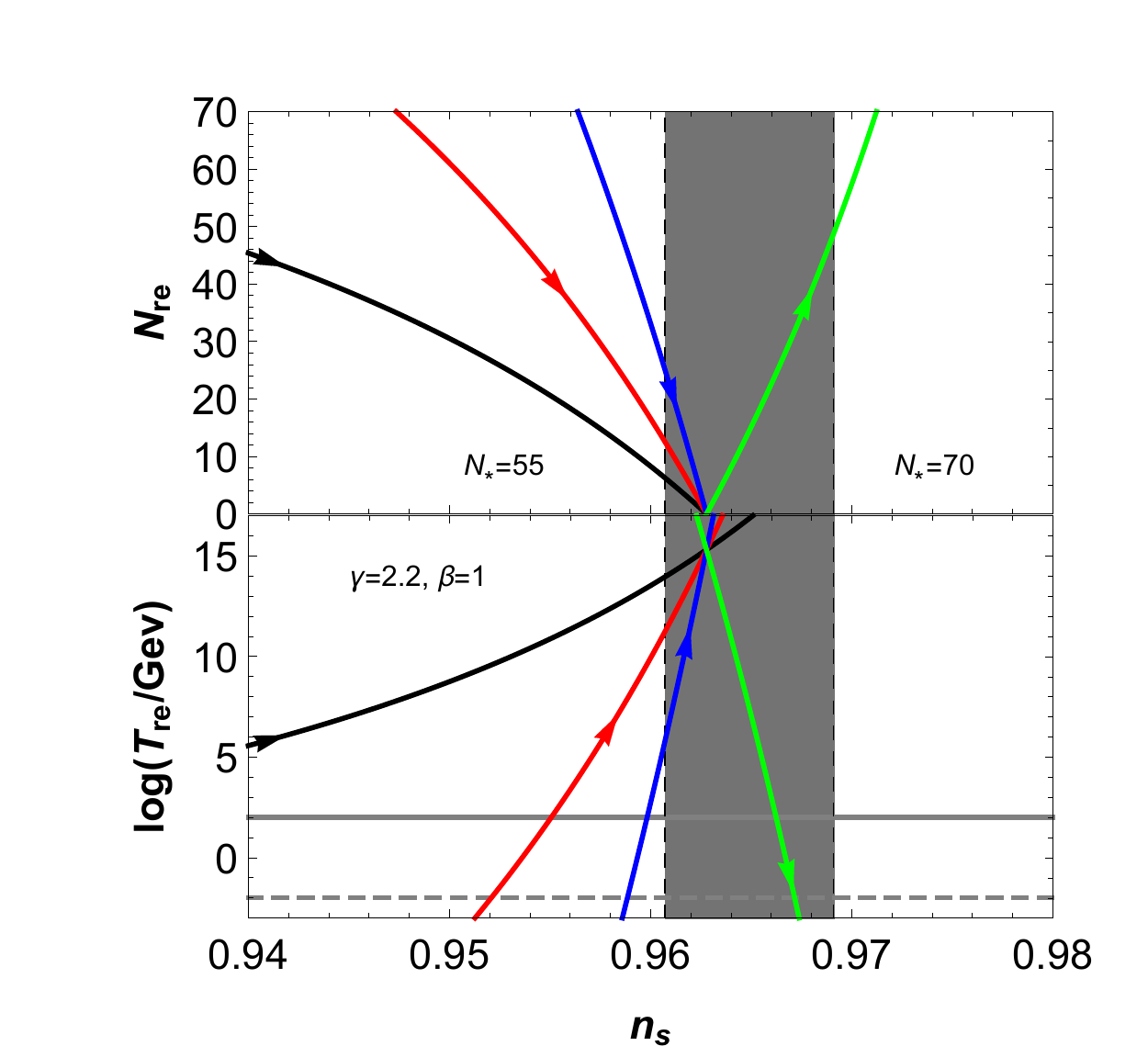}&
\includegraphics[width=0.45\textwidth]{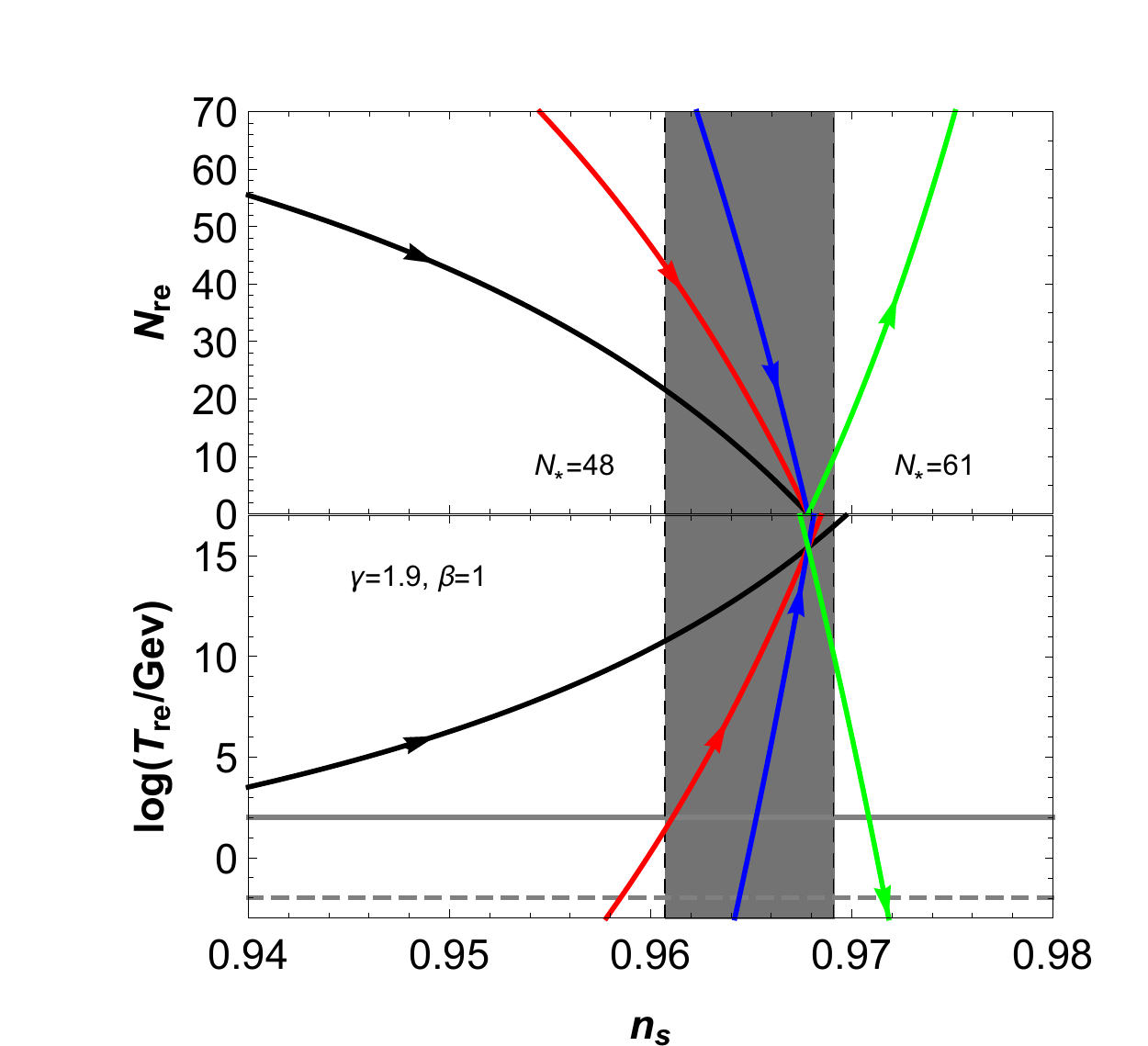}\\
\end{array}$
\caption{ (Top) The relations between $N_{re}$ and $n_s$; and (Bottom) the relations between $T_{re}$ and $n_s$. The~corresponding values of $\beta$ and $\gamma$ for each model are indicated in each panel.
The $1\sigma$ Planck constraint $n_s=0.9649\pm 0.0042$~\cite{Akrami:2018odb} is denoted by the gray band, and~the $1\sigma$ Planck constraint on the $e$-folds $N$ is also indicated.
The black, red, blue and green lines correspond to the reheating models with $w_{re}=-1/3$, 0, 1/6 and 2/3, respectively; in each line, the arrow denotes the direction of $N$ enlargement.
The horizontal gray solid and dashed lines in  the bottom  panels denote the electroweak scale $T_{EW}\sim 100$ GeV
and the big bang nucleosynthesis scale $T_{BBN}\sim 10$ MeV, respectively.}
\label{fig:reh2}
\end{figure}

For different kinds of $\beta$, $\gamma$, $N$ and $w_{re}$, by using Equations~\eqref{parar4}, \eqref{Nre:parar2}~and~\eqref{Tre:parar2}, we~calculate the corresponding spectral tilt $n_s$, reheating $e$-folds $N_{re}$ and reheating temperature $T_{re}$,
and the results are displayed in Figure~\ref{fig:reh2}.
The figures show that different model parameters $\beta$ and $\gamma$ and the value of $w_{re}$ provide different constraints on the reheating $e$-folds $N_{re}$ and the reheating temperature $T_{re}$, while the parameter $\beta$ almost does not affect the reheating process. For larger spectral tilt $n_s$, the allowed reheating $e$-folding number $N_{re}$ with $w_{re}=-1/3$, 0 and $1/6$ will become smaller,
while the allowed reheating $e$-folding number $N_{re}$ with $w_{re}=2/3$ will become larger. The~scale of big-bang nucleosynthesis put an upper  limit on $n_s$ if $w_{re}=2/3$ and a low limit on $n_s$ if $w_{re}=1/6$.

\section{Conclusions}\label{sec-5}
The non-minimal derivative coupling term in the inflation model could reduce the tensor-to-scalar ratio,  which can make the large tensor-to-scalar ratio models, such as the Higgs inflation, be consistent with the observations. We derive the reconstruction formulae of the inflation model with non-minimal derivative coupling. To reconstruct the potential without using the high friction limit, we~consider the parameterization of the tensor to scalar ratio $r=8\alpha/(N+\beta)^\gamma$ inspired from the $\alpha$-attractor. For $\gamma=2$, which is the $\alpha$ attractor, we~get the same potential as obtained in ~\cite{Yi:2016jqr}, in the GR limit $F\ll1$. When~$\alpha\ll1$, this potential has the same asymptotic behavior as that of
T/E-model. For~$\gamma\neq2$, the~potential is the exponential form. The~observational constraints on the parameters are $1.2<\gamma<2.7$ and $\beta<10$. The~reconstruction also show that the observational data favor the $\alpha$~attractor case with~$\gamma\sim2$.

The constraints on the spectral tilt $n_s$ from the Planck data could provide constraints on the reheating process. Different model parameters provide different constraints on reheating $e$-folds $N_{re}$, reheating temperature $T_{re}$ and reheating state equation $w_{re}$.
For larger spectral tilt $n_s$, the allowed reheating $e$-folding number $N_{re}$ with $w_{re}=-1/3$, 0 and $1/6$ will become smaller, while the allowed reheating $e$-folding number $N_{re}$ with $w_{re}=2/3$ will become larger.
The energy scale of the reheating could also provide additional constraints on the inflation. If $\gamma=2$, $\beta=1$ and $w_{re}=2/3$, the big bang nucleosynthesis scale requires $n_s<0.967$; if $\gamma=2$, $\beta=1$ and $w_{re}=1/6$, the big bang nucleosynthesis scale requires $n_s>0.962$.
\vspace{6pt}

\authorcontributions{Conceptualization, Z.Y.; investigation, Q.F.; data curation, Q.F.; writing---original draft preparation, Z.Y. and Y.Y.; and writing---review and editing, Z.Y. All authors have read and agreed to the published version of the manuscript.}

\funding{This research was supported in part by the National Natural Science Foundation of China under Grant No. 11947138, the Postdoctoral Science Foundation of China under Grant No. 2019M660514, the Hubei College Students' innovation and entrepreneurship training program under Grant No. S201910920050 and the Talent-Introduction Program of Hubei Polytechnic University under Grant No.19xjk25R.}

\acknowledgments{The authors thank Yungui Gong from Huazhong University of Science and Technology.}

\conflictsofinterest{The authors declare no conflict of interest.}

\reftitle{References}

\begin{thebibliography}{999}

\bibitem[Starobinsky(1980)]{Starobinsky:1980te}
Starobinsky, A.A.
\newblock {A New Type of Isotropic Cosmological Models Without Singularity}.
\newblock {\em Phys. Lett.} {\bf 1980}, {\em 91B}, 99--102,
\newblock
 doi:{\changeurlcolor{black}\href{https://doi.org/10.1016/0370-2693(80)90670-X}{\detokenize{10.1016/0370-2693(80)90670-X}}}.

\bibitem[Guth(1981)]{Guth:1980zm}
Guth, A.H.
\newblock {The Inflationary Universe: A Possible Solution to the Horizon and
 Flatness Problems}.
\newblock {\em Phys.~Rev.~D} {\bf 1981}, {\em 23}, 347--356,
\newblock
 doi:{\changeurlcolor{black}\href{https://doi.org/10.1103/PhysRevD.23.347}{\detokenize{10.1103/PhysRevD.23.347}}}.

\bibitem[Linde(1983)]{Linde:1983gd}
Linde, A.D.
\newblock {Chaotic Inflation}.
\newblock {\em Phys. Lett.} {\bf 1983}, {\em 129B}, 177--181,
\newblock
 doi:{\changeurlcolor{black}\href{https://doi.org/10.1016/0370-2693(83)90837-7}{\detokenize{10.1016/0370-2693(83)90837-7}}}.

\bibitem[Albrecht and Steinhardt(1982)]{Albrecht:1982wi}
Albrecht, A.; Steinhardt, P.J.
\newblock {Cosmology for Grand Unified Theories with Radiatively Induced
 Symmetry Breaking}.
\newblock {\em Phys. Rev. Lett.} {\bf 1982}, {\em 48}, 1220--1223,
\newblock
 doi:{\changeurlcolor{black}\href{https://doi.org/10.1103/PhysRevLett.48.1220}{\detokenize{10.1103/PhysRevLett.48.1220}}}.

\bibitem[Akrami \em{et al.}(2020)Akrami et al.]{Akrami:2018odb}
Akrami, Y.; Arroja, F.; Ashdown, M.; Aumont, J.; Baccigalupi, C.; Ballardini, M.; Banday, A.J.; Barreiro, R.B.; Bartolo, N.; Basak, S.; et al.
\newblock {Planck 2018 results. X. Constraints on inflation}.
\newblock {\em Astron. Astrophys.} {\bf 2020}, {\em 641}, A10,
\newblock
 doi:{\changeurlcolor{black}\href{https://doi.org/10.1051/0004-6361/201833887}{\detokenize{10.1051/0004-6361/201833887}}}.

\bibitem[Germani and Kehagias(2010)]{Germani:2010gm}
Germani, C.; Kehagias, A.
\newblock {New Model of Inflation with Non-minimal Derivative Coupling of
 Standard Model Higgs Boson to Gravity}.
\newblock {\em Phys. Rev. Lett.} {\bf 2010}, {\em 105}, 11302,
\newblock
 doi:{\changeurlcolor{black}\href{https://doi.org/10.1103/PhysRevLett.105.011302}{\detokenize{10.1103/PhysRevLett.105.011302}}}.

\bibitem[Germani \em{et al.}(2014)Germani, Watanabe, and
 Wintergerst]{Germani:2014hqa}
Germani, C.; Watanabe, Y.; Wintergerst, N.
\newblock {Self-unitarization of New Higgs Inflation and compatibility with
 Planck and BICEP2 data}.
\newblock {\em JCAP} {\bf 2014}, {\em 1412}, 9,
\newblock
 doi:{\changeurlcolor{black}\href{https://doi.org/10.1088/1475-7516/2014/12/009}{\detokenize{10.1088/1475-7516/2014/12/009}}}.

\bibitem[Horndeski(1974)]{Horndeski:1974wa}
Horndeski, G.W.
\newblock {Second-order scalar-tensor field equations in a four-dimensional
 space}.
\newblock {\em Int. J. Theor. Phys.} {\bf 1974}, {\em 10}, 363--384,
\newblock
 doi:{\changeurlcolor{black}\href{https://doi.org/10.1007/BF01807638}{\detokenize{10.1007/BF01807638}}}.

\bibitem[Sushkov(2009)]{Sushkov:2009hk}
Sushkov, S.V.
\newblock {Exact cosmological solutions with nonminimal derivative coupling}.
\newblock {\em Phys.~Rev.~D} {\bf 2009}, {\em 80}, 103505,
\newblock
 doi:{\changeurlcolor{black}\href{https://doi.org/10.1103/PhysRevD.80.103505}{\detokenize{10.1103/PhysRevD.80.103505}}}.

\bibitem[Yang \em{et al.}(2016)Yang, Fei, Gao, and~Gong]{Yang:2015pga}
Yang, N.; Fei, Q.; Gao, Q.; Gong, Y.~{Inflationary models with non-minimally derivative coupling}.
\newblock {\em Class.~Quant.~Grav.} {\bf 2016}, {\em 33}, 205001,
\newblock
 doi:{\changeurlcolor{black}\href{https://doi.org/10.1088/0264-9381/33/20/205001}{\detokenize{10.1088/0264-9381/33/20/205001}}}.

\bibitem[Yang \em{et al.}(2015)Yang, Gao, and~Gong]{Yang:2015zgh}
Yang, N.; Gao, Q.; Gong, Y.
\newblock {Inflation with non-minimally derivative coupling}.
\newblock {\em Int. J. Mod. Phys.} {\bf 2015}, {\em A30}, 1545004,
\newblock
 doi:{\changeurlcolor{black}\href{https://doi.org/10.1142/S0217751X15450049}{\detokenize{10.1142/S0217751X15450049}}}.

\bibitem[Huang \em{et al.}(2015)Huang, Gong, Liang, and~Yi]{Huang:2015yva}
Huang, Y.; Gong, Y.; Liang, D.; Yi, Z.
\newblock {Thermodynamics of scalar–tensor theory with non-minimally
 derivative coupling}.
\newblock {\em Eur. Phys. J. C} {\bf 2015}, {\em 75}, 351,
\newblock
 doi:{\changeurlcolor{black}\href{https://doi.org/10.1140/epjc/s10052-015-3574-7}{\detokenize{10.1140/epjc/s10052-015-3574-7}}}.

\bibitem[Gong \em{et al.}(2018)Gong, Papantonopoulos, and~Yi]{Gong:2017kim}
Gong, Y.; Papantonopoulos, E.; Yi, Z.~{Constraints on scalar–tensor theory of gravity by the recent observational results on gravitational waves}.~{\em Eur. Phys. J. C} {\bf 2018}, {\em 78}, 738,
\newblock
 doi:{\changeurlcolor{black}\href{https://doi.org/10.1140/epjc/s10052-018-6227-9}{\detokenize{10.1140/epjc/s10052-018-6227-9}}}.

\bibitem[Fu \em{et al.}(2019)Fu, Wu, and~Yu]{Fu:2019ttf}
Fu, C.; Wu, P.; Yu, H.
\newblock {Primordial Black Holes from Inflation with Nonminimal Derivative
 Coupling}.
\newblock {\em Phys.~Rev.~D} {\bf 2019}, {\em 100}, 63532,
\newblock
 doi:{\changeurlcolor{black}\href{https://doi.org/10.1103/PhysRevD.100.063532}{\detokenize{10.1103/PhysRevD.100.063532}}}.
\bibitem[Oikonomou and Fronimos(2020)]{Oikonomou:2020sij}
Oikonomou, V.; Fronimos, F.~Reviving Non-Minimal Horndeski-Like Theories after GW170817: Kinetic~ Coupling Corrected Einstein-Gauss-Bonnet Inflation. \emph{arXiv} {\bf 2020}, arXiv:gr-qc/2006.05512.

\bibitem[Odintsov \em{et al.}(2020)Odintsov, Oikonomou, and
 Fronimos]{Odintsov:2020sqy}
Odintsov, S.; Oikonomou, V.; Fronimos, F.
\newblock Rectifying Einstein-Gauss-Bonnet Inflation in View of GW170817.
\newblock {\em Nucl. Phys. B} {\bf 2020}, {\em 958}, 115135,
\newblock
 doi:{\changeurlcolor{black}\href{https://doi.org/10.1016/j.nuclphysb.2020.115135}{\detokenize{10.1016/j.nuclphysb.2020.115135}}}.

\bibitem[Gialamas \em{et al.}(2020)Gialamas, Karam, Lykkas, and
 Pappas]{Gialamas:2020vto}
Gialamas, I.D.; Karam, A.; Lykkas, A.; Pappas, T.D.
\newblock Palatini-Higgs inflation with nonminimal derivative coupling.
\newblock {\em Phys.~Rev.~D} {\bf 2020}, {\em 102}, 063522,
\newblock
 doi:{\changeurlcolor{black}\href{https://doi.org/10.1103/PhysRevD.102.063522}{\detokenize{10.1103/PhysRevD.102.063522}}}.


\bibitem[Kaiser(1995)]{Kaiser:1994vs}
Kaiser, D.I.
\newblock {Primordial spectral indices from generalized Einstein theories}.
\newblock {\em Phys.~Rev.~D} {\bf 1995}, {\em 52}, 4295--4306,
\newblock
 doi:{\changeurlcolor{black}\href{https://doi.org/10.1103/PhysRevD.52.4295}{\detokenize{10.1103/PhysRevD.52.4295}}}.

\bibitem[Bezrukov and Shaposhnikov(2008)]{Bezrukov:2007ep}
Bezrukov, F.L.; Shaposhnikov, M.
\newblock {The Standard Model Higgs boson as the inflaton}.
\newblock {\em Phys. Lett. B} {\bf 2008}, {\em 659}, 703--706,
\newblock
 doi:{\changeurlcolor{black}\href{https://doi.org/10.1016/j.physletb.2007.11.072}{\detokenize{10.1016/j.physletb.2007.11.072}}}.

\bibitem[Kallosh \em{et al.}(2013)Kallosh, Linde, and~Roest]{Kallosh:2013yoa}
Kallosh, R.; Linde, A.; Roest, D.~{Superconformal Inflationary $\alpha$-Attractors}.~{\em JHEP} {\bf 2013}, {\em 11}, 198,
\newblock
 doi:{\changeurlcolor{black}\href{https://doi.org/10.1007/JHEP11(2013)198}{\detokenize{10.1007/JHEP11(2013)198}}}.

\bibitem[Kallosh and Linde(2013{\natexlab{a}})]{Kallosh:2013hoa}
Kallosh, R.; Linde, A.~{Universality Class in Conformal Inflation}.~{\em JCAP} {\bf 2013}, {\em 1307}, 2,
\newblock
 doi:{\changeurlcolor{black}\href{https://doi.org/10.1088/1475-7516/2013/07/002}{\detokenize{10.1088/1475-7516/2013/07/002}}}.

\bibitem[Kallosh and Linde(2013{\natexlab{b}})]{Kallosh:2013maa}
Kallosh, R.; Linde, A.~{Non-minimal Inflationary Attractors}.~{\em JCAP} {\bf 2013}, {\em 1310}, 33,
\newblock
 doi:{\changeurlcolor{black}\href{https://doi.org/10.1088/1475-7516/2013/10/033}{\detokenize{10.1088/1475-7516/2013/10/033}}}.

\bibitem[Huang(2007)]{Huang:2007qz}
Huang, Q.G.
\newblock {Constraints on the spectral index for the inflation models in string
 landscape}.
\newblock {\em Phys.~Rev.~D} {\bf 2007}, {\em 76}, 61303,
\newblock
 doi:{\changeurlcolor{black}\href{https://doi.org/10.1103/PhysRevD.76.061303}{\detokenize{10.1103/PhysRevD.76.061303}}}.

\bibitem[Gobbetti \em{et al.}(2015)Gobbetti, Pajer, and
 Roest]{Gobbetti:2015cya}
Gobbetti, R.; Pajer, E.; Roest, D.~{On the Three Primordial Numbers}.~{\em JCAP} {\bf 2015}, {\em 1509}, 58,
\newblock
 doi:{\changeurlcolor{black}\href{https://doi.org/10.1088/1475-7516/2015/09/058}{\detokenize{10.1088/1475-7516/2015/09/058}}}.

\bibitem[Mukhanov(2013)]{Mukhanov:2013tua}
Mukhanov, V.
\newblock {Quantum Cosmological Perturbations: Predictions and Observations}.
\newblock {\em Eur. Phys. J. C} {\bf 2013}, {\em 73}, 2486,
\newblock
 doi:{\changeurlcolor{black}\href{https://doi.org/10.1140/epjc/s10052-013-2486-7}{\detokenize{10.1140/epjc/s10052-013-2486-7}}}.

\bibitem[Roest(2014)]{Roest:2013fha}
Roest, D.
\newblock {Universality classes of inflation}.
\newblock {\em JCAP} {\bf 2014}, {\em 1401}, 7,
\newblock
 doi:{\changeurlcolor{black}\href{https://doi.org/10.1088/1475-7516/2014/01/007}{\detokenize{10.1088/1475-7516/2014/01/007}}}.

\bibitem[Garcia-Bellido and Roest(2014)]{Garcia-Bellido:2014gna}
Garcia-Bellido, J.; Roest, D.
\newblock {Large-$N$ running of the spectral index of inflation}.
\newblock {\em Phys.~Rev.~D} {\bf 2014}, {\em 89}, 103527,
\newblock
 doi:{\changeurlcolor{black}\href{https://doi.org/10.1103/PhysRevD.89.103527}{\detokenize{10.1103/PhysRevD.89.103527}}}.

\bibitem[Garcia-Bellido \em{et al.}(2014{\natexlab{a}})Garcia-Bellido, Roest,
 Scalisi, and~Zavala]{Garcia-Bellido:2014wfa}
Garcia-Bellido, J.; Roest, D.; Scalisi, M.; Zavala, I.
\newblock {Lyth bound of inflation with a tilt}.
\newblock {\em Phys.~Rev.~D} {\bf 2014}, {\em 90}, 123539,
\newblock
 doi:{\changeurlcolor{black}\href{https://doi.org/10.1103/PhysRevD.90.123539}{\detokenize{10.1103/PhysRevD.90.123539}}}.

\bibitem[Garcia-Bellido \em{et al.}(2014{\natexlab{b}})Garcia-Bellido, Roest,
 Scalisi, and~Zavala]{Garcia-Bellido:2014eva}
Garcia-Bellido, J.; Roest, D.; Scalisi, M.; Zavala, I.
\newblock {Can CMB data constrain the inflationary field range?}
\newblock {\em JCAP} {\bf 2014}, {\em 1409}, 6,
\newblock
 doi:{\changeurlcolor{black}\href{https://doi.org/10.1088/1475-7516/2014/09/006}{\detokenize{10.1088/1475-7516/2014/09/006}}}.

\bibitem[Creminelli \em{et al.}(2015)Creminelli, Dubovsky, López Nacir,
 Simonović, Trevisan, Villadoro, and~Zaldarriaga]{Creminelli:2014nqa}
Creminelli, P.; Dubovsky, S.; López Nacir, D.; Simonović, M.; Trevisan, G.;
 Villadoro, G.; Zaldarriaga,~M.
\newblock {Implications of the scalar tilt for the tensor-to-scalar ratio}.~{\em Phys.~Rev.~D} {\bf 2015}, {\em 2}, 123528,
\newblock
 doi:{\changeurlcolor{black}\href{https://doi.org/10.1103/PhysRevD.92.123528}{\detokenize{10.1103/PhysRevD.92.123528}}}.

\bibitem[Boubekeur \em{et al.}(2015)Boubekeur, Giusarma, Mena, and
 Ramírez]{Boubekeur:2014xva}
Boubekeur, L.; Giusarma, E.; Mena, O.; Ramírez, H.
\newblock {Phenomenological approaches of inflation and their equivalence}.
\newblock {\em Phys.~Rev.~D} {\bf 2015}, {\em 91}, 83006,
\newblock
 doi:{\changeurlcolor{black}\href{https://doi.org/10.1103/PhysRevD.91.083006}{\detokenize{10.1103/PhysRevD.91.083006}}}.

\bibitem[Barranco \em{et al.}(2014)Barranco, Boubekeur, and
 Mena]{Barranco:2014ira}
Barranco, L.; Boubekeur, L.; Mena, O.
\newblock {A model-independent fit to Planck and BICEP2 data}.
\newblock {\em Phys.~Rev.~D} {\bf 2014}, {\em 90}, 63007,
\newblock
 doi:{\changeurlcolor{black}\href{https://doi.org/10.1103/PhysRevD.90.063007}{\detokenize{10.1103/PhysRevD.90.063007}}}.

\bibitem[Galante \em{et al.}(2015)Galante, Kallosh, Linde, and
 Roest]{Galante:2014ifa}
Galante, M.; Kallosh, R.; Linde, A.; Roest, D.
\newblock {Unity of Cosmological Inflation Attractors}.
\newblock {\em Phys. Rev. Lett.} {\bf 2015}, {\em 114}, 141302,
\newblock
 doi:{\changeurlcolor{black}\href{https://doi.org/10.1103/PhysRevLett.114.141302}{\detokenize{10.1103/PhysRevLett.114.141302}}}.

\bibitem[Chiba(2015)]{Chiba:2015zpa}
Chiba, T.~{Reconstructing the inflaton potential from the spectral index}.~{\em PTEP} {\bf 2015}, {\em 2015}, 73E02,
\newblock
 doi:{\changeurlcolor{black}\href{https://doi.org/10.1093/ptep/ptv090}{\detokenize{10.1093/ptep/ptv090}}}.

\bibitem[Cicciarella and Pieroni(2017)]{Cicciarella:2016dnv}
Cicciarella, F.; Pieroni, M.~{Universality for quintessence}.~{\em JCAP} {\bf 2017}, {\em 1708}, 10,
\newblock
 doi:{\changeurlcolor{black}\href{https://doi.org/10.1088/1475-7516/2017/08/010}{\detokenize{10.1088/1475-7516/2017/08/010}}}.

\bibitem[Lin \em{et al.}(2016)Lin, Gao, and~Gong]{Lin:2015fqa}
Lin, J.; Gao, Q.; Gong, Y.
\newblock {The reconstruction of inflationary potentials}.
\newblock {\em Mon. Not. Roy. Astron. Soc.} {\bf 2016}, {\em 459}, 4029--4037,
\newblock
 doi:{\changeurlcolor{black}\href{https://doi.org/10.1093/mnras/stw915}{\detokenize{10.1093/mnras/stw915}}}.

\bibitem[Nojiri and Odintsov(2011)]{Nojiri:2010wj}
Nojiri, S.; Odintsov, S.D.
\newblock {Unified cosmic history in modified gravity: from F(R) theory to
 Lorentz non-invariant models}.
\newblock {\em Phys. Rept.} {\bf 2011}, {\em 505}, 59--144,
\newblock
 doi:{\changeurlcolor{black}\href{https://doi.org/10.1016/j.physrep.2011.04.001}{\detokenize{10.1016/j.physrep.2011.04.001}}}.

\bibitem[Odintsov and Oikonomou(2016)]{Odintsov:2016vzz}
Odintsov, S.D.; Oikonomou, V.K.
\newblock {Inflationary $\alpha$-attractors from $F(R)$ gravity}.
\newblock {\em Phys.~Rev.~D} {\bf 2016}, {\em 94}, 124026,
\newblock
 doi:{\changeurlcolor{black}\href{https://doi.org/10.1103/PhysRevD.94.124026}{\detokenize{10.1103/PhysRevD.94.124026}}}.

\bibitem[Yi and Gong(2016)]{Yi:2016jqr}
Yi, Z.; Gong, Y.
\newblock {Nonminimal coupling and inflationary attractors}.
\newblock {\em Phys.~Rev.~D} {\bf 2016}, {\em 94}, 103527,
\newblock
 doi:{\changeurlcolor{black}\href{https://doi.org/10.1103/PhysRevD.94.103527}{\detokenize{10.1103/PhysRevD.94.103527}}}.

\bibitem[Odintsov and Oikonomou(2017)]{Odintsov:2017qpp}
Odintsov, S.D.; Oikonomou, V.K.
\newblock {Inflation with a Smooth Constant-Roll to Constant-Roll Era
 Transition}.
\newblock {\em Phys.~Rev.~D} {\bf 2017}, {\em 96}, 24029,
\newblock
 doi:{\changeurlcolor{black}\href{https://doi.org/10.1103/PhysRevD.96.024029}{\detokenize{10.1103/PhysRevD.96.024029}}}.

\bibitem[Nojiri \em{et al.}(2017)Nojiri, Odintsov, and
 Oikonomou]{Nojiri:2017qvx}
Nojiri, S.; Odintsov, S.D.; Oikonomou, V.K.
\newblock {Constant-roll Inflation in $F(R)$ Gravity}.
\newblock {\em Class. Quant. Grav.} {\bf 2017}, {\em 34}, 245012,
\newblock
 doi:{\changeurlcolor{black}\href{https://doi.org/10.1088/1361-6382/aa92a4}{\detokenize{10.1088/1361-6382/aa92a4}}}.

\bibitem[Choudhury(2017)]{Choudhury:2017cos}
Choudhury, S.~{COSMOS-$e'$- soft Higgsotic attractors}.~{\em Eur.~Phys.~J.~C} {\bf 2017}, {\em 77}, 469,
\newblock
 doi:{\changeurlcolor{black}\href{https://doi.org/10.1140/epjc/s10052-017-5001-8}{\detokenize{10.1140/epjc/s10052-017-5001-8}}}.

\bibitem[Gao and Gong(2018)]{Gao:2017uja}
Gao, Q.; Gong, Y.
\newblock {Reconstruction of extended inflationary potentials for attractors}.
\newblock {\em Eur. Phys. J. Plus} {\bf 2018}, {\em 133}, 491,
\newblock
 doi:{\changeurlcolor{black}\href{https://doi.org/10.1140/epjp/i2018-12324-3}{\detokenize{10.1140/epjp/i2018-12324-3}}}.

\bibitem[Jinno and Kaneta(2017)]{Jinno:2017jxc}
Jinno, R.; Kaneta, K.
\newblock {Hill-climbing inflation}.
\newblock {\em Phys.~Rev.~D} {\bf 2017}, {\em 96}, 43518,
\newblock
 doi:{\changeurlcolor{black}\href{https://doi.org/10.1103/PhysRevD.96.043518}{\detokenize{10.1103/PhysRevD.96.043518}}}.

\bibitem[Gao(2017)]{Gao:2017owg}
Gao, Q.
\newblock {Reconstruction of constant slow-roll inflation}.
\newblock {\em Sci. China Phys. Mech. Astron.} {\bf 2017}, {\em 60}, 90411,
\newblock
 doi:{\changeurlcolor{black}\href{https://doi.org/10.1007/s11433-017-9065-4}{\detokenize{10.1007/s11433-017-9065-4}}}.

\bibitem[Fei \em{et al.}(2017)Fei, Gong, Lin, and~Yi]{Fei:2017fub}
Fei, Q.; Gong, Y.; Lin, J.; Yi, Z.
\newblock {The reconstruction of tachyon inflationary potentials}.
\newblock {\em JCAP} {\bf 2017}, {\em 1708}, 18,
\newblock
 doi:{\changeurlcolor{black}\href{https://doi.org/10.1088/1475-7516/2017/08/018}{\detokenize{10.1088/1475-7516/2017/08/018}}}.

\bibitem[Koh \em{et al.}(2017)Koh, Lee, and~Tumurtushaa]{Koh:2016abf}
Koh, S.; Lee, B.H.; Tumurtushaa, G.
\newblock {Reconstruction of the Scalar Field Potential in Inflationary Models
 with a Gauss-Bonnet term}.
\newblock {\em Phys.~Rev.~D} {\bf 2017}, {\em 95}, 123509,
\newblock
 doi:{\changeurlcolor{black}\href{https://doi.org/10.1103/PhysRevD.95.123509}{\detokenize{10.1103/PhysRevD.95.123509}}}.

\bibitem[Dai \em{et al.}(2014)Dai, Kamionkowski, and~Wang]{Dai:2014jja}
Dai, L.; Kamionkowski, M.; Wang, J.
\newblock {Reheating constraints to inflationary models}.
\newblock {\em Phys. Rev. Lett.} {\bf 2014}, {\em 113}, 41302,
\newblock
 doi:{\changeurlcolor{black}\href{https://doi.org/10.1103/PhysRevLett.113.041302}{\detokenize{10.1103/PhysRevLett.113.041302}}}.

\bibitem[Cook \em{et al.}(2015)Cook, Dimastrogiovanni, Easson, and
 Krauss]{Cook:2015vqa}
Cook, J.L.; Dimastrogiovanni, E.; Easson, D.A.; Krauss, L.M.
\newblock {Reheating predictions in single field inflation}.
\newblock {\em JCAP} {\bf 2015}, {\em 1504}, 47,
\newblock
 doi:{\changeurlcolor{black}\href{https://doi.org/10.1088/1475-7516/2015/04/047}{\detokenize{10.1088/1475-7516/2015/04/047}}}.

\bibitem[Ueno and Yamamoto(2016)]{Ueno:2016dim}
Ueno, Y.; Yamamoto, K.
\newblock {Constraints on $\alpha$-attractor inflation and reheating}.
\newblock {\em Phys.~Rev.~D} {\bf 2016}, {\em 93}, 83524,
\newblock
 doi:{\changeurlcolor{black}\href{https://doi.org/10.1103/PhysRevD.93.083524}{\detokenize{10.1103/PhysRevD.93.083524}}}.

\bibitem[Kabir \em{et al.}(2016)Kabir, Mukherjee, and~Lohiya]{Kabir:2016kdh}
Kabir, R.; Mukherjee, A.; Lohiya, D.
\newblock {Reheating Constraints on K\"ahler Moduli Inflation.} \emph{arXiv} {\bf 2016}, arXiv:gr-qc/1609.09243.
\newblock 

\bibitem[Di Marco \em{et al.}(2017)Di Marco, Cabella, and
 Vittorio]{DiMarco:2017sqo}
Di Marco, A.; Cabella, P.; Vittorio, N.
\newblock {Reconstruction of $\alpha$-attractor supergravity models of
 inflation}.
\newblock {\em Phys.~Rev.~D} {\bf 2017}, {\em 95}, 23516,
\newblock
 doi:{\changeurlcolor{black}\href{https://doi.org/10.1103/PhysRevD.95.023516}{\detokenize{10.1103/PhysRevD.95.023516}}}.

\bibitem[Dimopoulos and Owen(2017)]{Dimopoulos:2017zvq}
Dimopoulos, K.; Owen, C.~{Quintessential Inflation with $\alpha$-attractors}.~{\em JCAP} {\bf 2017}, {\em 1706}, 27,
\newblock
 doi:{\changeurlcolor{black}\href{https://doi.org/10.1088/1475-7516/2017/06/027}{\detokenize{10.1088/1475-7516/2017/06/027}}}.

\bibitem[Gong \em{et al.}(2015)Gong, Pi, and~Leung]{Gong:2015qha}
Gong, J.O.; Pi, S.; Leung, G.~Probing reheating with primordial spectrum.~{\em JCAP} {\bf 2015}, {\em 5}, 27,
\newblock
 doi:{\changeurlcolor{black}\href{https://doi.org/10.1088/1475-7516/2015/05/027}{\detokenize{10.1088/1475-7516/2015/05/027}}}.

\bibitem[Tsujikawa(2012)]{Tsujikawa:2012mk}
Tsujikawa, S.
\newblock {Observational tests of inflation with a field derivative coupling to
 gravity}.
\newblock {\em Phys.~Rev.~D} {\bf 2012}, {\em 85}, 83518,
\newblock
 doi:{\changeurlcolor{black}\href{https://doi.org/10.1103/PhysRevD.85.083518}{\detokenize{10.1103/PhysRevD.85.083518}}}.

\end{thebibliography}

\end{document}